\documentclass[conference]{IEEEtran}
\usepackage[pdftex]{graphicx}
\usepackage{amsmath}
\usepackage{nicefrac}
\usepackage{balance}
\usepackage{url}
\usepackage{pgfplots}
\usepackage{multirow}
\usetikzlibrary{shapes.multipart,intersections}
\usepackage{cite}
\usepackage{caption}
\usepackage{amsmath,amssymb,amsfonts,steinmetz,bm}
\usepackage{mathtools,algpseudocode,algorithm,MnSymbol}
\usepackage{graphicx}
\usepackage{textcomp}
\usepackage{xcolor,comment}
\def\BibTeX{{\rm B\kern-.05em{\sc i\kern-.025em b}\kern-.08em
    T\kern-.1667em\lower.7ex\hbox{E}\kern-.125emX}}
\usepackage{gensymb}
\usepackage{multirow}
\usepackage{threeparttable}
\usepackage{hhline}
\usepackage{booktabs}
    
\usepackage{tikz}
\usetikzlibrary{arrows.meta}
\usetikzlibrary{calc}
\makeatletter
\newcommand{\gettikzxy}[3]{%
  \tikz@scan@one@point\pgfutil@firstofone#1\relax
  \edef#2{\the\pgf@x}%
  \edef#3{\the\pgf@y}%
}
\usepackage[nolist]{acronym}
\makeatother
\usepackage{upgreek} 
\usepackage{seqsplit}
\usepackage{balance}
\usepackage[font={footnotesize}]{caption}
\usepackage{subcaption}
\usepackage[inline]{enumitem}
\usepackage{amsthm}
\theoremstyle{plain}

\acrodef{AoD}{Angle of Departure}
\acrodef{AoA}{Angle of Arrival}
\acrodef{RIS}{Reconfigurable Intelligent Surface}
\acrodef{BS}{base station}
\acrodef{UE}{user equipment}
\acrodef{LoS}{line-of-sight}
\acrodef{NLoS}{non-line-of-sight}
\acrodef{NF}{nearfield}
\acrodef{SNR}{signal-to-noise ratio}
\acrodef{SINR}{signal-to-interference-and-noise-ratio}
\acrodef{SISO}{single-input-single-output}
\acrodef{FIM}{Fisher Information Matrix}
\acrodef{PEB}{position error bound}
\acrodef{SDP}{semidefinite program}
\acrodef{PSD}{positive semidefinite}
\acrodef{LMI}{linear matrix inequalities}
\acrodef{MC}{multi-carrier}
\acrodef{MIMO}{multiple inputs multiple outputs}
\acrodef{OEB}{orientation error bound}
\acrodef{DoD}{Direction of Departure}
\acrodef{TDoA}{Time Difference of Arrival}
\acrodef{w.r.t.}{with respect to}
\acrodef{SRE}{Smart radio environment}
\acrodef{TX}{transmitter}
\acrodef{RX}{receiver}
\acrodef{QoS}{Quality of Service}
\acrodef{VNA}{Vector Network Analyzer}
\acrodef{LS}{Least Squares}
\acrodef{(W)LS}{(Weighted) Least Squares}
\acrodef{WLS}{Weighted Least Squares}
\acrodef{RRIS}{Reflective RIS}
\acrodef{TRIS}{Transmit RIS}
\acrodef{MPC}{multipath component}
\acrodef{OTA}{Over-the-Air}
\acrodef{DP}{Direct Path}
\acrodef{RP}{Reflected Path}
\acrodef{SLAM}{Simultaneous Localization And Mapping} 
\acrodef{RMSE}{Root Mean Square Error}

\begin{document}
\title{RIS-aided Positioning Experiments based on mmWave Indoor Channel Measurements} 
\author{\IEEEauthorblockN{Moustafa Rahal\IEEEauthorrefmark{1}\IEEEauthorrefmark{3}, Beno\^{i}t Denis\IEEEauthorrefmark{1}, Taghrid Mazloum\IEEEauthorrefmark{1}, Frederic Munoz\IEEEauthorrefmark{1}, and Raffaele D'Errico\IEEEauthorrefmark{1}}
\IEEEauthorblockA{\IEEEauthorrefmark{1}CEA-Leti, Université Grenoble Alpes, F-38000 Grenoble, France\\
\IEEEauthorrefmark{3}Université Rennes 1, IETR - UMR 6164, F-35000 Rennes, France
}
}
\maketitle

\begin{abstract}
\acp{RIS} are announced as a truly transformative technology, capable of smartly shaping wireless environments to optimize next generation communication networks.  
Among their numerous foreseen applications, \acp{RRIS} have been shown theoretically beneficial not only to enable wireless localization through controlled multipath in situations where conventional systems would fail (e.g., with too few available \acp{BS} and/or under radio blockages) but also to locally boost accuracy on demand (typically, in regions close to the surface). In this paper, leveraging a dedicated frequency-domain mmWave indoor channel sounding campaign, we present the first experimental evidences of such benefits, by emulating offline simple \ac{RIS}-aided single-\ac{BS} positioning scenarios including \ac{LoS} and \ac{NLoS}, single-\ac{RIS} and multi-\ac{RIS}, and multiple \ac{UE} locations, also by considering various combinations of estimated multipath parameters (e.g., delays, \ac{AoD} or gains) as inputs to basic \ac{LS} solvers. Despite their simplicity, these preliminary proof-of-concept validations show concretely how and when \ac{RIS}-reflected paths could contribute to enhance localization performance.
\end{abstract}
\begin{IEEEkeywords}
Indoor Channel Sounding, Multipath Parameters, \ac{NLoS}, Proof-of-Concept Validations, \acl{RIS}, \acl{VNA}, Wireless Localization.
\end{IEEEkeywords}
\section{Introduction}
\acp{RIS}, which consist of controllable nearly-passive devices behaving as electromagnetic mirrors or lenses, are foreseen as an enabling breakthrough technology for beyond fifth generation (B5G) wireless systems~\cite{huang2019reconfigurable,qignqingwu2019}. These surfaces can purposely modify the wireless propagation environments to optimize communication networks in the sense of improved \ac{QoS}, extended coverage, low power consumption, limited field exposure \cite{RISE6G_COMMAG}, and more. Regarding wireless localization, \acp{RIS} have been shown not only to locally boost accuracy on demand, but also and foremost to enable localization feasibility in harsh operating contexts or under limited deployment settings for which conventional systems based on active \acp{BS} would fail~\cite{Kamran_Leveraging, WymeerschVehTechMag}. For instance, \acp{RRIS} have already been considered for parametric multipath-aided positioning in both \ac{LoS} (e.g., \cite{wymeersch_beyond_2020, Elzanaty2021, he_large_2020, zhang_towards_2020, keykhosravi2021siso}) or \ac{NLoS} (e.g., \cite{rahal2021ris, Rahal_Localization-Optimal_RIS, liu_reconfigurable_2020}) conditions, encompassing far and near field propagation regimes. However, most of these state-of-the-art contributions are still based on simplistic models and synthetic simulations for performance evaluation.

In contrast, this paper accounts for preliminary proof-of-concept validations of \ac{RIS}-aided positioning in a single-\ac{BS} indoor scenario, relying on frequency-domain mmWave channel measurement data. The main paper contributions can be summarized as follows: (i) we describe a dedicated \ac{VNA}-based channel measurement campaign, which was carried out with real \ac{RRIS} \cite{GNW_RRIS} and \ac{TRIS} \cite{CEA_TRIS} prototypes, both with beamforming capabilities, as well as a monopole antenna mounted on a positioner enabling high-resolution multipath estimation; (ii) we conduct an analysis of the \ac{RIS}-reflected multipath components in terms of delay, \ac{AoA} and \ac{AoD} (i.e., beyond overall channel gains) in light of localization needs; (iii) we benchmark positioning results for several \ac{UE} and \ac{RRIS} locations, while considering various combinations of estimated multipath parameters or overall channel parameters; (iv) we illustrate concrete limitations related to the presence of grating lobes at the \ac{RRIS} and/or too large distances between the \ac{RRIS} and the \ac{UE}. To the best of our knowledge, these experiments represent a world premiere demonstrating the validity of the \ac{RIS}-aided localization concept at mmWave frequencies based on a real-life surface prototype.  

\begin{figure}[t]
 \centering
 \includegraphics[width=1\linewidth]{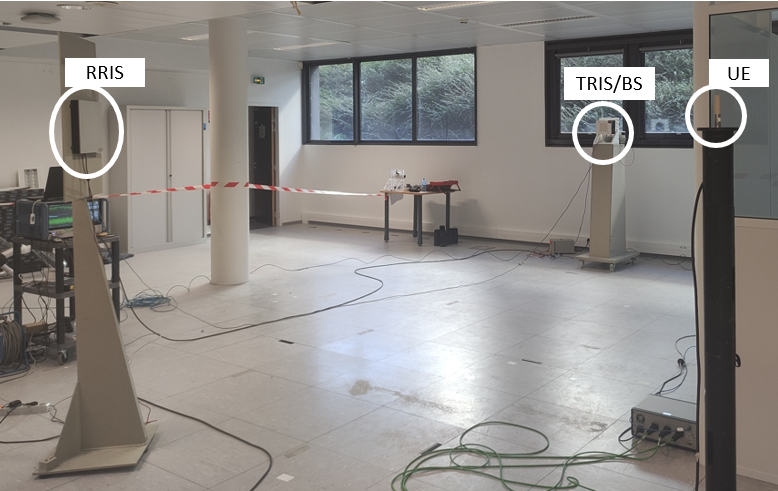}
 \caption{Picture of the indoor environment and equipment considered for the mmWave measurement campaign.}
 \vspace{-5mm}
 \label{fig:picture}
 \end{figure}
 \begin{figure*}[t]
 \centering
 \includegraphics[width=0.75\linewidth]{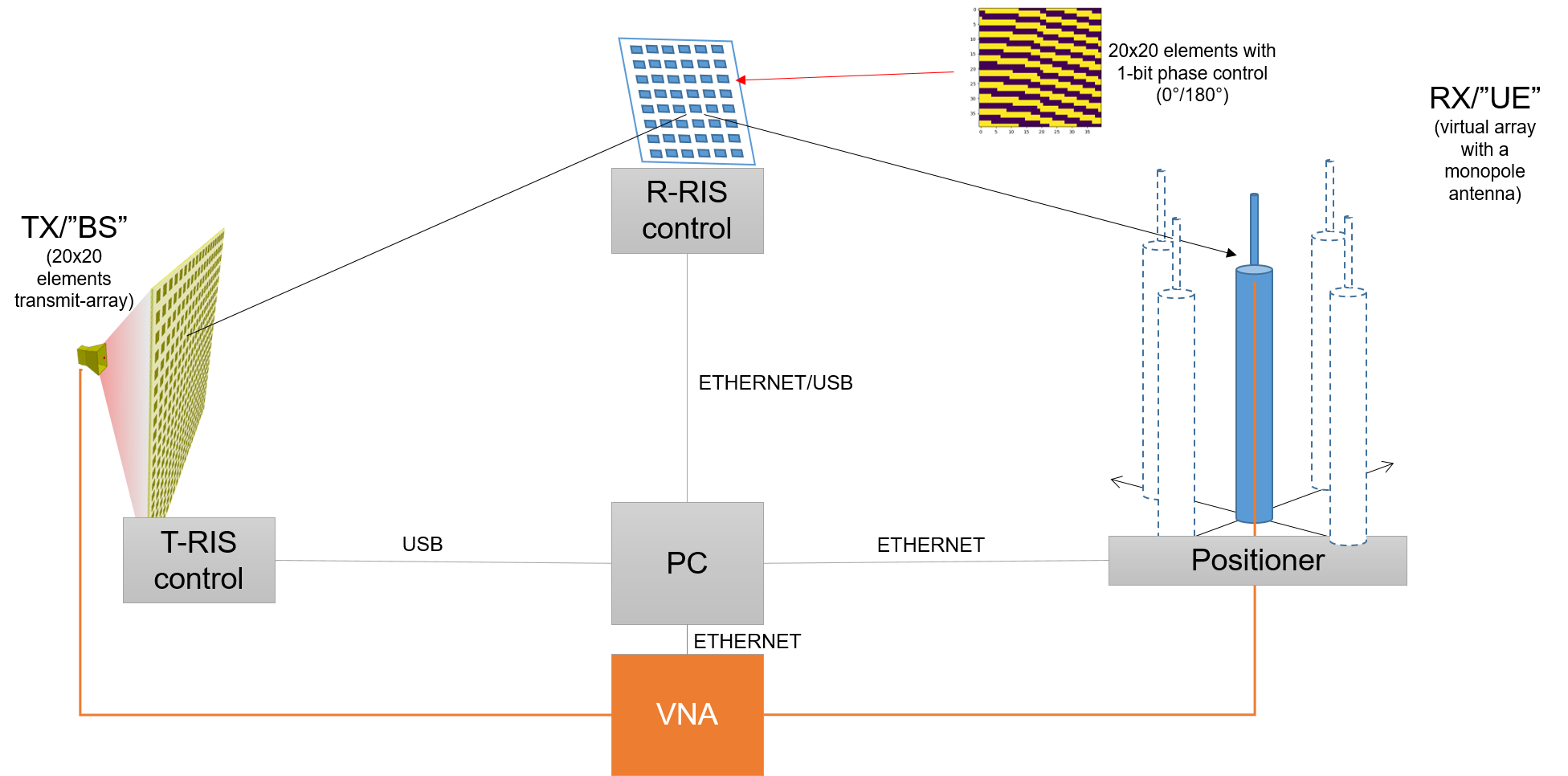}
 \caption{Simplified block diagram of the \ac{RIS}-enabled mmWave \ac{VNA}-based channel sounder, with 1-bit \ac{RRIS} phase control.}
 \vspace{-5mm}
 \label{fig:system}
 \end{figure*}
\begin{figure}[t]
 \centering
 \includegraphics[width=0.9\linewidth]{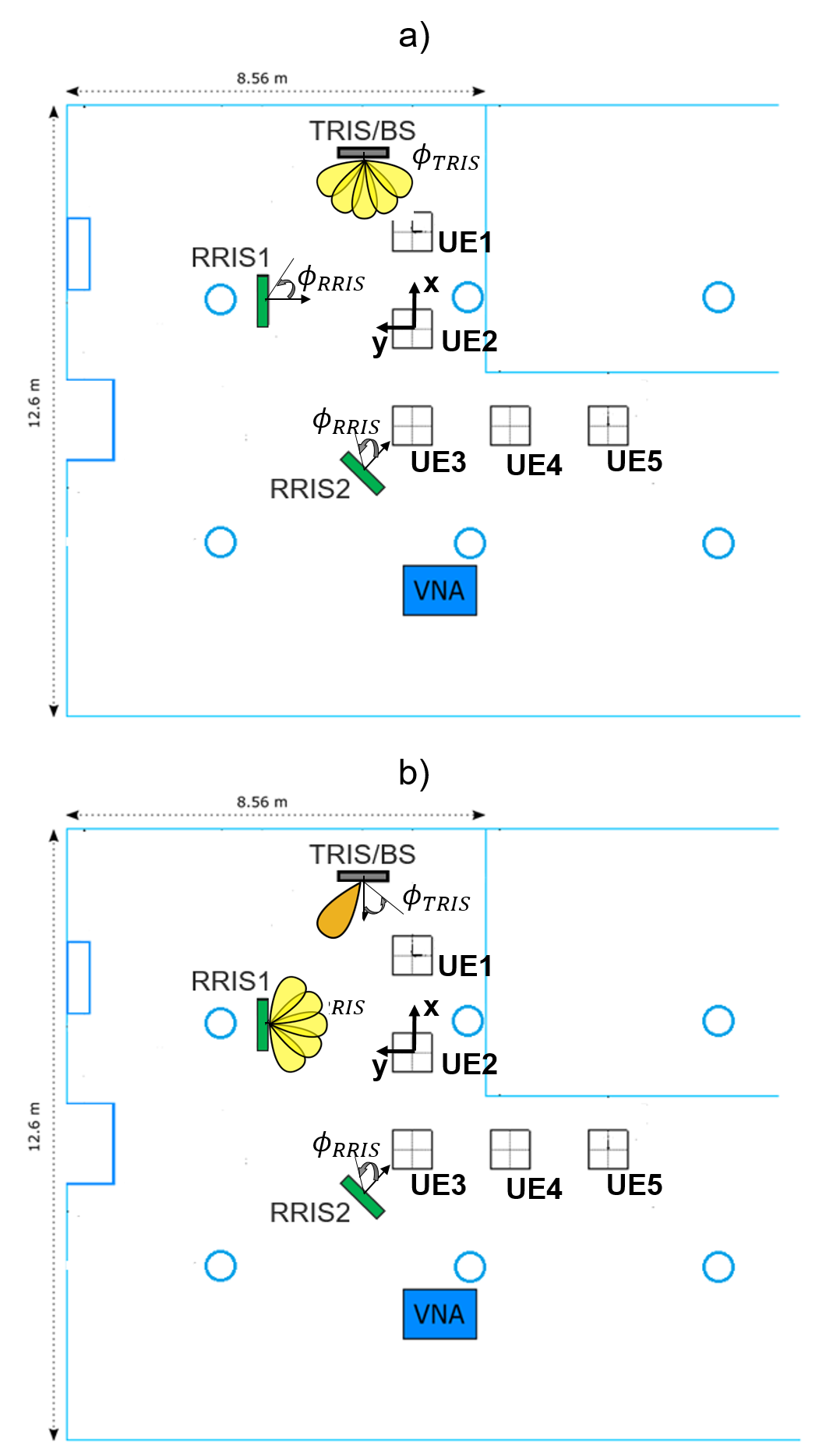}
 \caption{Layout and deployment considered in the mmWave channel measurement campaign, including 1 \ac{BS} location, 2 \ac{RRIS} locations (RRIS1 and 2) and 5 \ac{UE} locations (UE1 to UE5), represented for both the \ac{BS} and RRIS1 beams sweeping phases (resp. a and b).}
 \vspace{-5mm}
 \label{fig:layout}
 \end{figure}

\section{mmWave Channel Measurements}\label{sec:measurements}

\subsection{Experimental Setup and Measurement Procedure}\label{subsec:measurement_setup}

The experimental setup committed in our measurement campaign relies on a VNA-based mmWave channel sounder, which is similar to that used in \cite{Channel_EuCNC23}. It includes a \ac{TRIS} \cite{CEA_TRIS} on the \ac{TX} side and a \ac{RRIS} \cite{GNW_RRIS} between the \ac{TX} and the \ac{RX}, both with 1-bit element-wise phase control. In the following descriptions, unless specified, we reserve the term "\ac{RIS}" to the reflective \ac{RIS}, while systematically depicting the \ac{TRIS} as "\ac{BS}" for the sake of simplicity. Fig.~\ref{fig:system} shows a simplified block diagram of the overall measurement acquisition chain. Both the floor plan of the reference indoor environment and the tested deployment configurations are represented in Fig.~\ref{fig:layout}. 
Lying in a unique reference location for all the tested configurations, the BS first performed a beam scanning 
in azimuth from $-60\degree$ to $+60\degree$, by a step of $5\degree$, while the \ac{RRIS} was still off. Then, using a static beam, the \ac{BS} just illuminated the activated \ac{RRIS}, which was tested sequentially within two distinct location/orientation configurations. For each location/orientation setting, the \ac{RRIS} was controlled through a codebook to perform also beam sweeping in azimuth from $-60\degree$ to $+60\degree$, still by the same step of $5\degree$. For the scanning process of both \ac{BS} and \ac{RRIS}, a unique anticlockwise convention was used to define the \acp{AoD}, using a reference angle (i.e., $0\degree$) normal to the surface/array. The measurement procedure above was repeated over 5 main \ac{UE} positions. In each of those positions, the \ac{RX} monopole antenna was moved over a $3\times3$ small-scale grid (i.e., considering a virtual square array in the horizontal $x$-$y$ plane) thanks to a high-precision positioner, and a frequency-domain complex channel response between the \ac{TX} and the \ac{RX} was recorded from $25$ to $35$ GHz 
by the step of $10$ MHz on each occupied small-scale position, although a bandwidth of $2$ GHz (among the measured $10$ GHz bandwidth) was further used in practice in our localization tests (See Sec.~\ref{sec:preprocessing}) to emulate the behaviour of a realistic receiver. Note that the three involved entities (i.e., \ac{BS}, \ac{RRIS} and \ac{UE}) were all set at the same height of $1.6$m and hence, lying on the same 2D plane during all experiments. 

\subsection{Measurement Data Calibration and Pre-processing}\label{sec:preprocessing}

\begin{figure}[t]
 \centering
 \resizebox{\columnwidth}{!}{\input{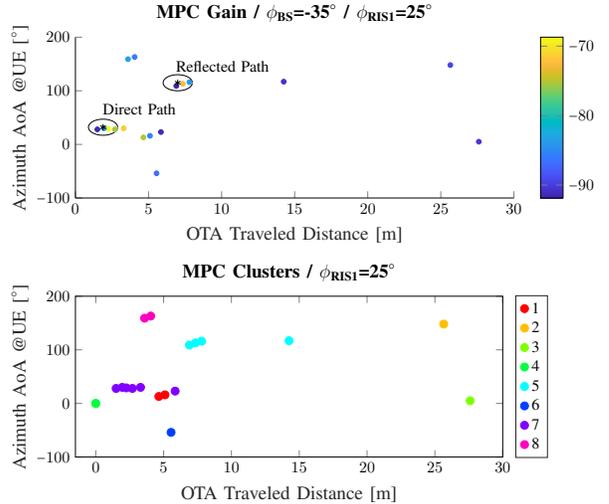}}
 \caption{Example of scatter plot of extracted \ac{MPC} gains (in dB) as a function of both their \acp{AoA} (in $\degree$) and traveled \ac{OTA} distances (in m) (top), along with the corresponding \ac{MPC} clusters (bottom), for the \ac{BS} beam pointing to the first \ac{RRIS} location (i.e., $\phi_{BS}=-35 \degree $) and the same \ac{RRIS} pointing to the first \ac{UE} location (i.e., $\phi_{RIS1}=25 \degree $).}
 \vspace{-5mm}
 \label{fig:ex_SAGE_extract}
 \end{figure}

The effect of all cables and RF components in the acquisition chain was pre-characterized and calibrated out of all the recorded complex frequency-domain channel responses. Then, thanks to the virtual $3\times3$ array used at the \ac{RX} \ac{UE}, for each tested location and pointing beam direction (at the \ac{BS} or the \ac{RRIS}), the classical high-resolution Space-Alternating Generalized Expectation-maximization (SAGE) algorithm \cite{SAGE} was applied to extract the parameters of the most significant \acp{MPC}, including the travelled \ac{OTA} distance (or equivalently, the delay\footnote{As we operate with a \ac{VNA}, \ac{TX}-\ac{RX} synchronization is inherently solved and any estimated delay directly coincides with the absolute time of flight, after calibration. In a real asynchronous system relying on relative delay estimation at the \ac{RX} though, either multi-way protocol exchanges, or joint localization and synchronization algorithms \cite{keykhosravi2021siso}, would be necessary.}), and the \ac{AoA} and the gain. For this extraction, as a stopping rule, we have considered retrieving up to 99\% of the total channel energy, with a limitation to the 20 strongest \acp{MPC}. SAGE was applied to a $2$GHz sub-band centered around the \ac{RRIS} operating frequency, that is, between $27$ and $29$ GHz. In the following analysis, among all the \acp{MPC} extracted by SAGE, both the \ac{BS}-\ac{UE} direct path (whenever available) and the \ac{BS}-\ac{RRIS}-\ac{UE} reflected path were isolated\footnote{Focusing mostly on effects such as signal-to-noise ratio or geometry in our analysis, the detection of a so-called \emph{isolated} \ac{MPC} is herein idealized for simplicity. It is genie-aided in the sense that we are interested only in the \acp{MPC} extracted in a distance-\ac{AoA} region where the geometric path (i.e., the \ac{DP} or a \ac{RIS} \ac{RP}) is expected to lie within an arbitrary tolerance margin of $\pm 1m$ in terms of travelled \ac{OTA} distance and $\pm 15\degree$ in terms of \ac{AoA} (See the black circles in Fig.~\ref{fig:ex_SAGE_extract}). In turn, more advanced processing would be necessary to isolate those contributions (typically, after subtracting all the static contributions from non-\ac{RIS} \acp{MPC} out of channel responses), while in a real dynamic system, further processing such as filtering could be applied to leverage prior information about the \ac{UE} location.} using the expected geometric information in terms of travelled distance and \ac{AoA} (i.e., performing genie-aided space-time filtering), while the overall channel gain was computed out of the gains of all the resolved \acp{MPC}.

\section{Radio Features and Positioning}\label{sec:features_and_positioning}

\subsection{Location-dependent Radio Features}\label{subsec:radio_features}

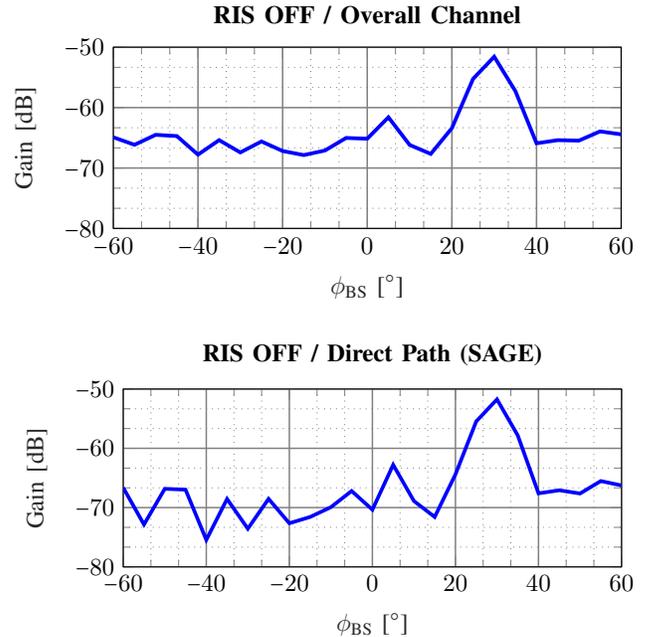
\begin{figure}[t]
  \centering
  \begin{subfigure}[b]{\columnwidth}
         \centering
        \resizebox{1\columnwidth}{!}{
%
%
\definecolor{mycolor1}{rgb}{0.00000,0.44700,0.74100}%
\begin{tikzpicture}

\begin{axis}[%
width=2.8in,
height=1in,
at={(0.758in,0.752in)},
scale only axis,
xmin=-60,
xmax=60,
xlabel style={font=\normalsize \color{white!15!black}},
xlabel={$\phi{}_{\text{BS}}\text{ [}^\circ\text{]}$},
ymin=-80,
ymax=-50,
ylabel style={font=\normalsize \color{white!15!black}},
ylabel={Gain [dB]},
axis background/.style={fill=white},
title style={font=\normalsize \bfseries},
title={RIS OFF / Overall Channel},
minor tick num=2,
grid = both,
major grid style={line width=.6pt,draw=black!50},
minor grid style={draw=black!50, dotted},
]
\addplot [color=blue, line width=1.5pt, forget plot]
  table[row sep=crcr]{%
-60	-64.9117099285811\\
-55	-66.1561578074424\\
-50	-64.5052341203864\\
-45	-64.7063764576707\\
-40	-67.7830653285155\\
-35	-65.3848997396845\\
-30	-67.421548288116\\
-25	-65.6038997703818\\
-20	-67.1813423790886\\
-15	-67.8588519820689\\
-10	-67.1026013173452\\
-5	-65.0214110790303\\
0	-65.1499997704991\\
5	-61.5982104070789\\
10	-66.1686461635838\\
15	-67.6656281217742\\
20	-63.3947222134523\\
25	-55.2461914687788\\
30	-51.5703840285143\\
35	-57.260429539393\\
40	-65.9086654002125\\
45	-65.3882551468269\\
50	-65.4653659340538\\
55	-63.9388135078118\\
60	-64.4258489230172\\
};
\end{axis}

\begin{axis}[%
width=2.8in,
height=1.4in,
at={(0in,0in)},
scale only axis,
xmin=0,
xmax=1,
ymin=0,
ymax=1,
axis line style={draw=none},
ticks=none,
axis x line*=bottom,
axis y line*=left
]
\end{axis}
\end{tikzpicture}%
}
     \end{subfigure}
            \vskip\baselineskip \vspace{-8mm}
       \begin{subfigure}[b]{\columnwidth}
         \centering
        \resizebox{1\columnwidth}{!}{
%
%
%
\begin{tikzpicture}

\begin{axis}[%
width=2.8in,
height=1in,
at={(0.828in,0.752in)},
scale only axis,
xmin=-60,
xmax=60,
xlabel style={font=\normalsize \color{white!15!black}},
xlabel={$\phi{}_{\text{BS}}\text{ [}^\circ\text{]}$},
xminorticks=true,
ymin=-80,
ymax=-50,
ylabel style={font=\normalsize \color{white!15!black}},
ylabel={Gain [dB]},
yminorticks=true,
axis background/.style={fill=white},
title style={font=\normalsize \bfseries},
title={RIS OFF / Direct Path (SAGE)},
minor tick num=2,
grid = both,
major grid style={line width=.6pt,draw=black!50},
minor grid style={draw=black!50, dotted},
]
\addplot [color=blue, line width=1.5pt, forget plot]
  table[row sep=crcr]{%
-60	-66.6435773666539\\
-55	-72.8544632986689\\
-50	-66.8331572771577\\
-45	-66.9725815045831\\
-40	-75.4519420853523\\
-35	-68.5473492998604\\
-30	-73.564876913759\\
-25	-68.5249847245232\\
-20	-72.6447183183387\\
-15	-71.5839403827195\\
-10	-69.9340426406617\\
-5	-67.1982725697994\\
0	-70.3599428282743\\
5	-62.7866707824396\\
10	-68.8955053442353\\
15	-71.5809028167459\\
20	-64.3902559343072\\
25	-55.4640782109165\\
30	-51.7433704194662\\
35	-57.8135952974943\\
40	-67.5976502754891\\
45	-67.0789351546758\\
50	-67.6235061683714\\
55	-65.5207986124153\\
60	-66.2674792101105\\
};
\end{axis}

\begin{axis}[%
width=2.8in,
height=1.4in,
at={(0in,0in)},
scale only axis,
xmin=0,
xmax=1,
ymin=0,
ymax=1,
axis line style={draw=none},
ticks=none,
axis x line*=bottom,
axis y line*=left
]
\end{axis}
\end{tikzpicture}%
}
     \end{subfigure}
     \vspace{-8mm}
  \caption{Overall channel (top) and \ac{DP} (bottom) gains in the first tested location (\ac{UE}1) after \ac{BS} beam scanning, as a function of $\phi_{BS}$, with the \ac{RRIS} off.}
 \vspace{-3mm}
\label{fig:ex_BS_scan}
\end{figure}

\begin{figure}[t]
 \centering
 \resizebox{1\columnwidth}{!}{
%
%
%
\begin{tikzpicture}

\begin{axis}[%
width=1.858in,
height=1.17in,
at={(0.853in,2.633in)},
scale only axis,
xmin=-60,
xmax=60,
xlabel style={font=\Large \color{white!15!black}},
xlabel={$\phi{}_{\text{RIS1}}\text{ [}^\circ\text{]}$},
ymin=-65.4625726593309,
ymax=-64.6462598798014,
ylabel style={font=\large \color{white!15!black}},
ylabel={Gain [dB]},
axis background/.style={fill=white},
title style={font=\bfseries, align=center},
title={RIS1 ON / Overall Channel\\ $\phi{}_{\text{BS}}\text{=-35}^\circ$},
minor tick num=2,
grid = both,
major grid style={line width=.6pt,draw=black!50},
minor grid style={draw=black!50, dotted},
]
\addplot [color=blue, line width=1.5pt, forget plot]
  table[row sep=crcr]{%
-60	-65.3370175804534\\
-55	-65.3377525049305\\
-50	-65.3397006627373\\
-45	-65.3525381477191\\
-40	-65.3175689754551\\
-35	-65.3172309507052\\
-30	-65.3328439727983\\
-25	-65.2859611184517\\
-20	-65.3610734393938\\
-15	-65.3563918675114\\
-10	-65.3218005403915\\
-5	-65.3374098539122\\
0	-64.9005885817389\\
5	-65.1164095418643\\
10	-65.2979397592376\\
15	-65.3658772266748\\
20	-65.2658838285465\\
25	-64.6462598798014\\
30	-64.942911656344\\
35	-65.4332553871642\\
40	-65.4460909492072\\
45	-65.431824247627\\
50	-65.4625726593309\\
55	-65.4583424933898\\
60	-65.3614024279995\\
};
\addplot [color=black, line width = 1.5, dashed, forget plot]
  table[row sep=crcr]{%
25	-65.4625726593309\\
25	-64.6462598798014\\
};
\addplot [color=black, line width = 1.5,dashed, forget plot]
  table[row sep=crcr]{%
0	-65.4625726593309\\
0	-64.6462598798014\\
};
\addplot [color=black,line width = 1.5, dashed, forget plot]
  table[row sep=crcr]{%
30	-65.4625726593309\\
30	-64.6462598798014\\
};
\addplot [color=red, line width = 1.5,forget plot]
  table[row sep=crcr]{%
24.58	-65.4625726593309\\
24.58	-64.6462598798014\\
};
\end{axis}

\begin{axis}[%
width=1.858in,
height=1.17in,
at={(0.853in,0.56in)},
scale only axis,
xmin=-60,
xmax=60,
xlabel style={font=\Large \color{white!15!black}},
xlabel={$\phi{}_{\text{RIS1}}\text{ [}^\circ\text{]}$},
ymin=-100,
ymax=-72.7228382858527,
ylabel style={font=\large \color{white!15!black}},
ylabel={Gain [dB]},
axis background/.style={fill=white},
title style={font=\bfseries, align=center},
title={RIS1 ON / Reflected Path (SAGE)\\$\phi{}_{\text{BS}}\text{=-35}^\circ$},
minor tick num=2,
grid = both,
major grid style={line width=.6pt,draw=black!50},
yshift=-1cm,
minor grid style={draw=black!50, dotted},
]
\addplot [color=blue, line width=1.5pt, forget plot]
  table[row sep=crcr]{%
-60	-92.2332943960524\\
-55	-93.1912578419029\\
-50	-92.7834601240397\\
-45	-100\\
-40	-86.4490394028049\\
-35	-86.707462772233\\
-30	-87.4784111396436\\
-25	-86.1957096365413\\
-20	-90.2159276390359\\
-15	-88.4295715112804\\
-10	-83.1276059271697\\
-5	-84.5741473144506\\
0	-74.6045867268289\\
5	-76.9252864487692\\
10	-83.4652113094625\\
15	-86.3031944765613\\
20	-80.1812247417728\\
25	-72.7228382858527\\
30	-74.8772380624712\\
35	-90.9416856556111\\
40	-87.9907638871376\\
45	-86.1206559504228\\
50	-100\\
55	-90.381772706257\\
60	-83.0827719387846\\
};
\addplot [color=black, dashed, line width = 1.5, forget plot]
  table[row sep=crcr]{%
25	-100\\
25	-72.7228382858527\\
};
\addplot [color=black, dashed, line width = 1.5, forget plot]
  table[row sep=crcr]{%
0	-100\\
0	-72.7228382858527\\
};
\addplot [color=black, dashed, line width = 1.5, forget plot]
  table[row sep=crcr]{%
30	-100\\
30	-72.7228382858527\\
};
\addplot [color=red,line width = 1.5, forget plot]
  table[row sep=crcr]{%
24.58	-100\\
24.58	-72.7228382858527\\
};
\end{axis}

\begin{axis}[%
width=1.858in,
height=1.17in,
at={(3.422in,2.633in)},
scale only axis,
xmin=-60,
xmax=60,
xlabel style={font=\Large \color{white!15!black}},
xlabel={$\phi{}_{\text{RIS2}}\text{ [}^\circ\text{]}$},
ymin=-64.8682500075994,
ymax=-64.6697171900174,
ylabel style={font=\large \color{white!15!black}},
ylabel={ },
axis background/.style={fill=white},
title style={font=\bfseries, align=center},
title={RIS2 ON / Overall Channel\\ $\phi{}_{\text{BS}}\text{=0}^\circ$},
minor tick num=2,
grid = both,
major grid style={line width=.6pt,draw=black!50},
minor grid style={draw=black!50, dotted},
]
\addplot [color=blue, line width=1.5pt, forget plot]
  table[row sep=crcr]{%
-60	-64.8378904058264\\
-55	-64.8333469543456\\
-50	-64.8458383811277\\
-45	-64.8366315261294\\
-40	-64.8499586348485\\
-35	-64.849127467793\\
-30	-64.8577633623727\\
-25	-64.842294862555\\
-20	-64.846599986623\\
-15	-64.8622829504081\\
-10	-64.8682500075994\\
-5	-64.841334762232\\
0	-64.8189831234273\\
5	-64.7987151292785\\
10	-64.7686704577055\\
15	-64.8356781542648\\
20	-64.8055890933424\\
25	-64.8215238140438\\
30	-64.7753900295617\\
35	-64.6697171900174\\
40	-64.7725550289029\\
45	-64.773335532794\\
50	-64.78419458754\\
55	-64.7603480677766\\
60	-64.7778081547761\\
};
\addplot [color=black, dashed, line width = 1.5, forget plot]
  table[row sep=crcr]{%
35	-64.8682500075994\\
35	-64.6697171900174\\
};
\addplot [color=black, dashed, line width = 1.5, forget plot]
  table[row sep=crcr]{%
55	-64.8682500075994\\
55	-64.6697171900174\\
};
\addplot [color=black, dashed, line width = 1.5, forget plot]
  table[row sep=crcr]{%
10	-64.8682500075994\\
10	-64.6697171900174\\
};
\addplot [color=red, line width = 1.5, forget plot]
  table[row sep=crcr]{%
31.91	-64.8682500075994\\
31.91	-64.6697171900174\\
};
\end{axis}

\begin{axis}[%
width=1.858in,
height=1.17in,
at={(3.422in,0.56in)},
scale only axis,
xmin=-60,
xmax=60,
xlabel style={font=\Large \color{white!15!black}},
xlabel={$\phi{}_{\text{RIS2}}\text{ [}^\circ\text{]}$},
ymin=-100,
ymax=-74.6045867268289,
ylabel style={font=\large \color{white!15!black}},
ylabel={ },
axis background/.style={fill=white},
title style={font=\bfseries, align=center},
title={RIS2 ON / Reflected Path (SAGE)\\$\phi{}_{\text{BS}}\text{=0}^\circ$},
minor tick num=2,
grid = both,
major grid style={line width=.6pt,draw=black!50},
yshift=-1cm,
minor grid style={draw=black!50, dotted},
]
\addplot [color=blue, line width=1.5pt, forget plot]
  table[row sep=crcr]{%
-60	-92.2332943960524\\
-55	-93.1912578419029\\
-50	-92.7834601240397\\
-45	-100\\
-40	-86.4490394028049\\
-35	-86.707462772233\\
-30	-87.4784111396436\\
-25	-87.6651264699597\\
-20	-90.2159276390359\\
-15	-87.4070143330948\\
-10	-83.1276059271697\\
-5	-91.3729907728074\\
0	-74.6045867268289\\
5	-82.9259818595377\\
10	-80.7974779689225\\
15	-90.6347088007448\\
20	-89.5586042141593\\
25	-89.5024374028905\\
30	-85.2992143051654\\
35	-79.825456320279\\
40	-85.1134921641002\\
45	-89.8265729906998\\
50	-100\\
55	-90.381772706257\\
60	-92.217389025415\\
};
\addplot [color=black, dashed, line width = 1.5, forget plot]
  table[row sep=crcr]{%
0	-100\\
0	-74.6045867268289\\
};
\addplot [color=black, dashed, line width = 1.5, forget plot]
  table[row sep=crcr]{%
35	-100\\
35	-74.6045867268289\\
};
\addplot [color=black, dashed, line width = 1.5, forget plot]
  table[row sep=crcr]{%
10	-100\\
10	-74.6045867268289\\
};
\addplot [color=red, line width = 1.5, forget plot]
  table[row sep=crcr]{%
31.91	-100\\
31.91	-74.6045867268289\\
};
\end{axis}

\begin{axis}[%
width=1.8in,
height=1.17in,
at={(0in,0in)},
scale only axis,
xmin=0,
xmax=1,
ymin=0,
ymax=1,
axis line style={draw=none},
ticks=none,
axis x line*=bottom,
axis y line*=left
]
\end{axis}
\end{tikzpicture}%
}
 \caption{Overall channel (top) and \ac{RP} (bottom) gains in first position (\ac{UE}) after \ac{RRIS} beam scanning, as a function of $\phi_{RIS1}$ with $\phi_{BS}=-35\degree$ (left) or $\phi_{RIS2}$ with $\phi_{BS}=0\degree$ (right), with its Ground-Truth angle (solid red) and the corresponding 3 strongest candidate estimates (dashed black)
 }
 \vspace{-5mm}
 \label{fig:ex_RIS_scan}
 \end{figure}
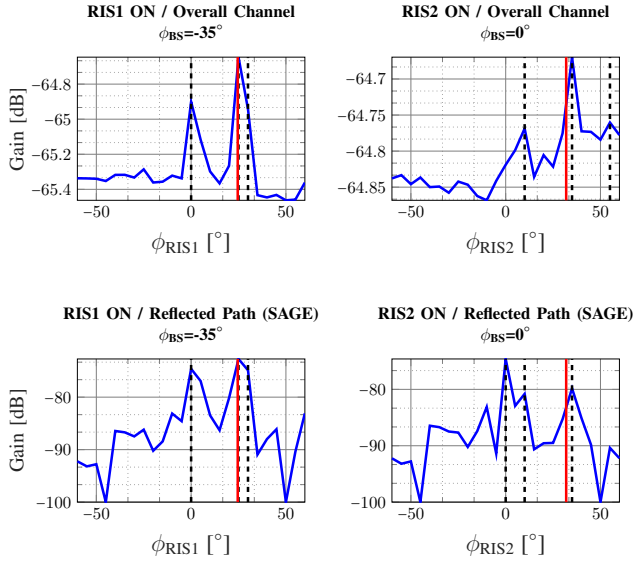

Picking up the maximum values taken by the overall channel gain over all possible beam pointing directions (i.e., in azimuth, discretized by steps of $5\degree$), we first come up with coarse \ac{AoD} estimates, $\widetilde{\phi}_{BS}$ and $\widetilde{\phi}_{RIS}$, respectively after \ac{BS} and \ac{RRIS} beam sweepings. Likewise, based on \acp{MPC} extraction, we also consider fine \ac{AoD} estimates, based on the three maximum values taken by the gains of isolated \acp{MPC} (i.e., the \ac{DP} or a \ac{RIS} \ac{RP}) over all possible pointing directions, after performing \ac{BS} and \ac{RRIS} beam sweeping. Fig.~\ref{fig:ex_BS_scan} and \ref{fig:ex_RIS_scan} show examples of such \ac{AoD} estimation with respect to the first \ac{UE} location (i.e., UE1), respectively for \ac{DP} after \ac{BS} beam scanning and for \ac{RP} after beam \ac{RRIS} scanning (for two distinct \ac{RRIS} locations). In this illustration, $\widetilde{\phi}_{BS} = 30.0\degree$ for a ground-truth angle of $32.0\degree$ (See Fig.~\ref{fig:ex_BS_scan}), while $\widetilde{\phi}_{RIS1} = 25\degree/0\degree/30\degree$ for a ground-truth angle of $24.6\degree$ (See Fig.~\ref{fig:ex_RIS_scan} - left) and $\widetilde{\phi}_{RIS2} = 0\degree/35\degree/10\degree$ for a ground-truth angle of $31.9\degree$ (See Fig.~\ref{fig:ex_RIS_scan} - right).

Besides, calibrated delays (or equivalently, \ac{OTA} traveled distances) associated with the same isolated \acp{MPC} have also been considered to complete angular estimates in some positioning scenarios (See Sec.~\ref{subsec:positioning_scenarios}). 

Finally, estimated \acp{AoA} are herein used for spatially filtering out the \acp{MPC} of interest, although they could be used also for \ac{UE} orientation estimation, which falls out of the scope of this paper.

\subsection{Positioning Scenarios and Method}\label{subsec:positioning_scenarios}

In all the tested positioning configurations, we assume that both the positions and the orientations of \ac{BS} and/or \ac{RRIS} are perfectly known. Table~\ref{Table:Positioning_Scenarios} summarizes the corresponding scenarios, depending on the combination of location-dependent radio features extracted from channel responses, as well as the expected benefits from using \ac{RIS} on top a conventional single-\ac{BS} system (i.e., either enabling or boosting localization). 

In a real system, from a protocol standpoint, the scenarios making use of several \acp{MPC} (i.e., all except 0, 2d and 2e) would necessitate a sequential illumination sequence from the \ac{BS} (e.g., \ac{BS} beam scanning first whenever \ac{DP} is used and/or \acp{RIS} beam scanning under static \ac{BS} beam), coming up with one set of estimated location-dependent radio parameters for each \ac{BS} pointing direction, which need to be recombined offline for positioning. In all these multipath scenarios, for the purpose of fair comparisons and putting more emphasis on the impact of radio parameters estimation, positioning is systematically determined through standard non-linear \ac{LS} optimization, initialized with the same random guess (drawn in the entire tested indoor area) and fed by the best candidate \ac{AoD} angles. In a real system, the latter optimistic assumption could be relaxed by leveraging prior knowledge of the \ac{RIS} beam patterns (as a function of the impinging angle) and/or the output of a tracking filter in case of dynamic scenario (i.e., both previous \ac{UE} estimates and its related uncertainty). 

Other positioning approaches could have been considered too, such as that relying on weighted combinations of the received powers (or power gains in our case) over a grid of possible \ac{UE} candidate locations (e.g., \cite{Kamran_Leveraging}), hence guaranteeing more homogeneous 2D spatial resolution regardless to the \ac{RIS}-\ac{UE} distance. However, the latter method would have required the prior discretization of the 2D Cartesian space during measurements collection and the pre-calculation of \ac{BS} and \ac{RRIS} beams accordingly, rather than the simpler azimuth domain discretization imposed here.     

\begin{table*}
\renewcommand{\arraystretch}{1.5}
\centering
\scriptsize
\caption{Positioning scenarios as a a function of combined radio metrics.}
\begin{threeparttable}
\begin{tabular}{|c|c|c|c|c|}
\hline
\textbf{Scenario} & \textbf{MPCs [Nb R-RISs]} & \textbf{LoS/NLoS Status} & \textbf{Radio Metrics [Method]} & \textbf{Expected RIS Benefits} \\ \hhline{|=====|}
0 & DP only [0 R-RIS] & LoS & $\widetilde{\phi}_{BS}$, $\widetilde{d}_{DP}$ [SAGE MPCs] & N/A (Baseline) \\ \hline 
1a & DP + RP1 [1 R-RIS] & LoS & $\widetilde{\phi}_{BS}$, $\widetilde{\phi}_{RIS1}$ [Overall Channel Gain] & Enabled localization\tnote{*} \\ \hline
1b & DP + RP2 [1 R-RIS] & LoS & $\widetilde{\phi}_{BS}$, $\widetilde{\phi}_{RIS2}$ [Overall Channel Gain] & Enabled localization\tnote{*} \\ \hline
1c & DP + RP1 + RP2 [2 R-RISs] & LoS & $\widetilde{\phi}_{BS}$, $\widetilde{\phi}_{RIS1}$, $\widetilde{\phi}_{RIS2}$ [Overall Channel Gain] & Enabled localization\tnote{*} \\ \hline
1d & RP1 + RP2 [2 R-RISs] & NLoS & $\widetilde{\phi}_{RIS1}$, $\widetilde{\phi}_{RIS2}$ [Overall Channel Gain] & Enabled localization\tnote{*} \\ \hline
2a & DP + RP1 [1 R-RIS] & LoS & $\widetilde{\phi}_{BS}$, $\widetilde{d}_{DP}$, $\widetilde{\phi}_{RIS1}$, $\widetilde{d}_{RP1}$ [SAGE MPCs] & Boosted localization\tnote{**} \\ \hline
2b & DP + RP2 [1 R-RIS] & LoS & $\widetilde{\phi}_{BS}$, $\widetilde{d}_{DP}$, $\widetilde{\phi}_{RIS2}$, $\widetilde{d}_{RP2}$ [SAGE MPCs] & Boosted localization\tnote{**} \\ \hline
2c & DP + RP1 + RP2 [2 R-RISs] & LoS & $\widetilde{\phi}_{BS}$, $\widetilde{d}_{DP}$, $\widetilde{\phi}_{RIS1}$, $\widetilde{d}_{RP1}$,$\widetilde{\phi}_{RIS2}$, $\widetilde{d}_{RP2}$ [SAGE MPCs] & Boosted localization\tnote{**} \\ \hline
2d & RP1 [1 R-RIS] & NLoS & $\widetilde{\phi}_{RIS1}$, $\widetilde{d}_{RP1}$ [SAGE MPCs] & Enabled localization\tnote{***} \\ \hline
2e & RP2 [1 R-RIS] & NLoS & $\widetilde{\phi}_{RIS2}$, $\widetilde{d}_{RP2}$ [SAGE MPCs] & Enabled localization\tnote{***} \\ \hline
2f & RP1 + RP2 [2 R-RISs] & NLoS & $\widetilde{\phi}_{RIS1}$, $\widetilde{d}_{RP1}$, $\widetilde{\phi}_{RIS2}$, $\widetilde{d}_{RP2}$ [SAGE MPCs] & Enabled localization\tnote{***} \\ \hline
\end{tabular}
\begin{tablenotes}
       \item[*] Vs. conventional single-BS positioning using also RSS measurements (i.e., missing additional links wrt. other BSs).
       \item[**] Vs. conventional single-BS positioning with similar MPCs estimation capabilities, but relying on DP only (See Scenario 0). 
       \item[***] Vs. conventional single-BS positioning with similar MPCs estimation capabilities, but relying on DP only (i.e., missing additional \ac{LoS} links wrt. additional BSs), or with no extra \ac{SLAM} capabilities (i.e., unable to position scatterers out of the extracted MPCs, jointly with UE).
\end{tablenotes}
\end{threeparttable} \label{Table:Positioning_Scenarios}
\end{table*}

\section{Results}\label{sec:results}
\subsection{Radio Parameters Estimation}\label{subsec:multipath_parameters_estimation}
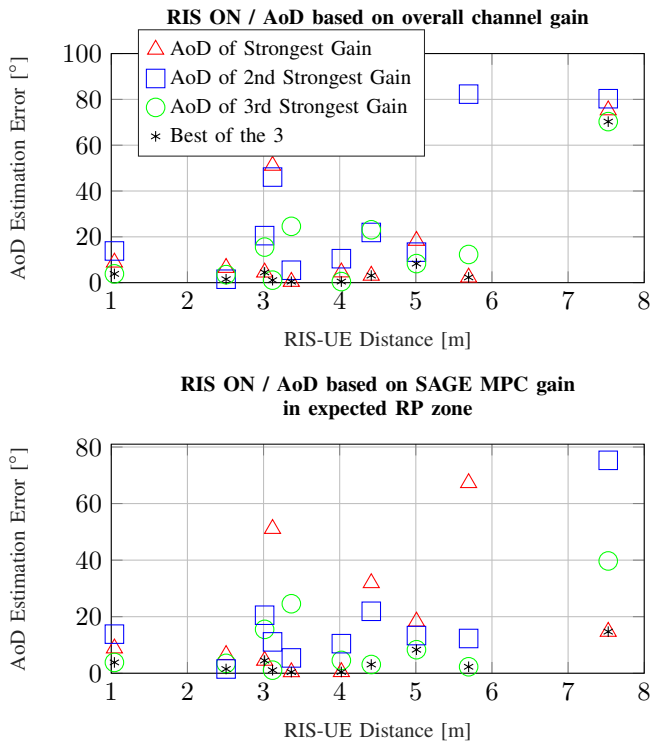
\begin{figure}[t]
 \centering
 {
%
%
\begin{tikzpicture}

\begin{axis}[%
width=0.8\columnwidth,
height=1.2in,
at={(2.08in,4.554in)},
scale only axis,
xmin=1,
xmax=8,
xlabel style={font= \footnotesize \color{white!15!black}},
xlabel={RIS-UE Distance [m]},
ymin=0,
ymax=100,
ylabel style={font= \footnotesize \color{white!15!black}},
ylabel={$\text{AoD Estimation Error [}^\circ\text{]}$},
axis background/.style={fill=white},
title style={font= \footnotesize \bfseries , align=center},
title={RIS ON / AoD based on overall channel gain},
yshift = -1.65in,
legend style={font=\footnotesize, at={(0.05,0.55)},  anchor=south west, legend cell align=left, align=left, draw=white!15!black},
grid = both,
minor grid style={line width=.005pt,draw=black!50, dashed},
]
\addplot [color=black, only marks, mark size=3.5pt, mark=triangle, mark options={solid, red}]
  table[row sep=crcr]{%
4.41474801092883	3.09\\
2.50798724079689	6.5\\
1.04120122935002	8.83\\
3.01398407427777	4.48\\
5.00840293906151	18.32\\
3.36505572019246	0.420000000000002\\
3.11826875044471	51.09\\
4.02190253487078	4.54000000000001\\
5.69347872570013	2.29000000000001\\
7.52699807360146	75.29\\
};
\addlegendentry{AoD of Strongest Gain}

\addplot [color=black, only marks, mark size=3.5pt, mark=square, mark options={solid, blue}]
  table[row sep=crcr]{%
3.36505572019246	5.42\\
3.11826875044471	46.09\\
4.02190253487078	10.46\\
5.69347872570013	82.29\\
7.52699807360146	80.29\\
4.41474801092883	21.91\\
2.50798724079689	1.5\\
1.04120122935002	13.83\\
3.01398407427777	20.52\\
5.00840293906151	13.32\\
};
\addlegendentry{AoD of 2nd Strongest Gain}

\addplot [color=black, only marks, mark size=3.5pt, mark=o, mark options={solid, green}]
  table[row sep=crcr]{%
3.36505572019246	24.58\\
3.11826875044471	1.09\\
4.02190253487078	0.460000000000008\\
5.69347872570013	12.29\\
7.52699807360146	70.29\\
4.41474801092883	23.09\\
2.50798724079689	3.5\\
1.04120122935002	3.83\\
3.01398407427777	15.52\\
5.00840293906151	8.32\\
};
\addlegendentry{AoD of 3rd Strongest Gain}

\addplot [color=black, only marks, mark=asterisk, mark options={solid, black}]
  table[row sep=crcr]{%
4.41474801092882	3.09\\
2.50798724079689	1.5\\
1.04120122935002	3.83\\
3.01398407427777	4.48\\
5.00840293906151	8.32\\
3.36505572019246	0.420000000000002\\
3.11826875044471	1.09\\
4.02190253487078	0.460000000000008\\
5.69347872570013	2.29000000000001\\
7.52699807360146	70.29\\
};
\addlegendentry{Best of the 3}
\end{axis}

\begin{axis}[%
width=0.8\columnwidth,
height=1.2in,
at={(2.08in,0.858in)},
scale only axis,
xmin=1,
xmax=8,
xlabel style={font= \footnotesize \color{white!15!black}},
xlabel={RIS-UE Distance [m]},
ymin=0,
ymax=81,
ylabel style={font= \footnotesize \color{white!15!black}},
ylabel={$\text{AoD Estimation Error [}^\circ\text{]}$},
axis background/.style={fill=white},
title style={font= \footnotesize \bfseries, align=center},
title={RIS ON / AoD based on SAGE MPC gain\\in expected RP zone},
legend style={font=\footnotesize, at={(0.331,0.45)},  anchor=south west, legend cell align=left, align=left, draw=white!15!black},
grid = both,
minor grid style={line width=.005pt,draw=black!50, dashed},
]
\addplot [color=black, only marks, mark size=3.5pt, mark=triangle, mark options={solid, red}]
  table[row sep=crcr]{%
3.36505572019246	0.420000000000002\\
3.11826875044471	51.09\\
4.02190253487078	0.460000000000008\\
5.69347872570013	67.29\\
7.52699807360146	14.71\\
4.41474801092883	31.91\\
2.50798724079689	6.5\\
1.04120122935002	8.83\\
3.01398407427777	4.48\\
5.00840293906151	18.32\\
};

\addplot [color=black, only marks, mark size=3.5pt, mark=o, mark options={solid, green}]
  table[row sep=crcr]{%
3.36505572019246	24.58\\
3.11826875044471	1.09\\
4.02190253487078	4.54\\
5.69347872570013	2.29\\
7.52699807360145	39.71\\
4.41474801092882	3.09\\
2.50798724079689	3.5\\
1.04120122935002	3.83\\
3.01398407427777	15.52\\
5.00840293906151	8.32\\
};

\addplot [color=black, only marks, mark size=3.5pt, mark=square, mark options={solid, blue}]
  table[row sep=crcr]{%
3.36505572019246	5.42\\
3.11826875044471	11.09\\
4.02190253487078	10.46\\
5.69347872570013	12.29\\
7.52699807360146	75.29\\
4.41474801092883	21.91\\
2.50798724079689	1.5\\
1.04120122935002	13.83\\
3.01398407427777	20.52\\
5.00840293906151	13.32\\
};

\addplot [color=black, only marks, mark=asterisk, mark options={solid, black}]
  table[row sep=crcr]{%
3.36505572019246	0.420000000000002\\
3.11826875044471	1.09\\
4.02190253487078	0.460000000000001\\
5.69347872570013	2.29\\
7.52699807360145	14.71\\
4.41474801092882	3.09\\
2.50798724079689	1.5\\
1.04120122935002	3.83\\
3.01398407427777	4.48\\
5.00840293906151	8.32\\
};

\end{axis}

\end{tikzpicture}%

}
 \caption{\ac{RIS} \ac{AoD} estimation errors, as a function of the true \ac{RIS}-\ac{UE} distance, relying on 3 maximum values taken by overall channel gain (top) or SAGE \acp{MPC} gain (bottom), over all \ac{RRIS} and \ac{UE} locations.}
 \label{fig:AoD_estim_errors}
 \end{figure}

\begin{figure}[t]
 \centering
 \resizebox{1\columnwidth}{!}
 {
%
%
\begin{tikzpicture}

\begin{axis}[%
width=4.521in,
height=1.476in,
at={(0.758in,2.571in)},
scale only axis,
xmin=0,
xmax=8,
xlabel style={font=\large \color{white!15!black}},
xlabel={RIS-UE Traveled Distance [m]},
ymin=0,
ymax=0.5,
ylabel style={font=\large \color{white!15!black}},
ylabel={Distance Estimation Error [m]},
axis background/.style={fill=white},
title style={font= \large \bfseries},
title={RIS ON / Distance based on SAGE MPC in Expected RP Zone},
legend style={legend cell align=left, align=left, draw=white!15!black},
grid = both,
minor grid style={line width=.005pt,draw=black!50, dashed},
]
\addplot [color=black, only marks, mark=triangle, mark size = 4.5pt, mark options={solid, black}]
  table[row sep=crcr]{%
3.36505572019246	0.0500000000000007\\
3.11826875044471	0.140000000000001\\
4.02190253487078	0.14\\
5.69347872570013	0.119999999999999\\
7.52699807360145	0.0800000000000001\\
};
\addlegendentry{RP1}

\addplot [color=black, mark size = 4.5pt, only marks, mark=asterisk, mark options={solid, black}]
  table[row sep=crcr]{%
4.41474801092882	0.19\\
2.50798724079689	0.290000000000001\\
1.04120122935002	0.11\\
3.01398407427777	0.24\\
5.00840293906151	0.19\\
};
\addlegendentry{RP2}

\end{axis}
\end{tikzpicture}%
}
 \caption{Multipath distance estimation errors, as a function of the true \ac{RIS}-\ac{UE} distance for reflected paths RP1 and RP2, over all \ac{RRIS} and \ac{UE} locations.}
 \label{fig:Dist_estim_errors}
 \end{figure}
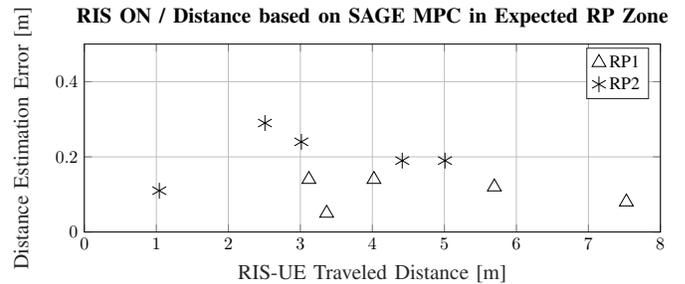

Fig.~\ref{fig:AoD_estim_errors} and \ref{fig:Dist_estim_errors} 
show 
estimation errors respectively for \ac{RIS} \acp{AoD} and the overall \ac{OTA} distances travelled by \ac{RIS}-reflected paths (i.e., from \ac{BS} to \ac{UE}, through the \ac{RRIS}), both 
as a function of the true \ac{RIS}-\ac{UE} distance (i.e., over the 5 \ac{UE} locations and the 2 \ac{RIS} locations). 
For \acp{AoD} first, we show the errors corresponding to the 3 maximum values taken by overall channel gains (top) or SAGE \ac{MPC} gains (bottom) over all possible azimuth angles, along with the best choice (among these 3 candidates).
As expected, one can observe 
that the \ac{AoD} error globally increases 
with the \ac{RIS}-\ac{UE} distance, staying typically below $5\degree$ (i.e., within the azimuth discretization step) at distances up to 5m but growing to several tens of degrees for distances beyond 5m, even in the best case. Whatever the \ac{RIS} location, the same trends are globally observed with the two estimation methods, which anyway lead roughly to the same estimates in a majority of tested cases. Nevertheless, the use of \ac{MPC} gains seems to outperform that of overall channel gains at larger \ac{RIS}-\ac{UE} distances (typically with errors up to $14.7\degree$ and $70.3\degree$, respectively, at $7.5$m from the first \ac{RIS}). This is likely due to the fact that, at larger \ac{RIS}-\ac{UE} distances, the dynamics of the overall channel gain as a function of \ac{RIS} \ac{AoD} is relatively limited and hence, more challenging to interpret. Both approaches suffer similarly from the presence of strong grating lobes in the \ac{RIS} beam patterns, which tends to generate local maxima (at wrong \ac{AoD} angles, possibly very distant from the ground-truth) in both channel and reflected path gains, even at short distances (See Fig.~\ref{fig:ex_RIS_scan}).
It shall be also noted that the angular step of 5° applied for beam scanning is quite large compared to the \ac{RRIS} -3dB beamwidth (i.e.,  below 3° for the considered prototype \cite{GNW_RRIS}). This effect contributes to the fact that some grating lobes could contribute even more than the apparent main lobe. Accordingly, the angle leading to the strongest power gains does not even always coincide with the best candidate.  
%
Regarding the distance estimation of \ac{RIS}-reflected paths, on Fig.~\ref{fig:Dist_estim_errors}, the 
negative influence of the \ac{RIS}-\ac{UE} distance looks less obvious than for \ac{AoD} estimation. The latter is thus expected to be the dominating factor with respect to positioning performance in our evaluations.

%
%

\subsection{2D Position Estimation}\label{subsec:positioning}

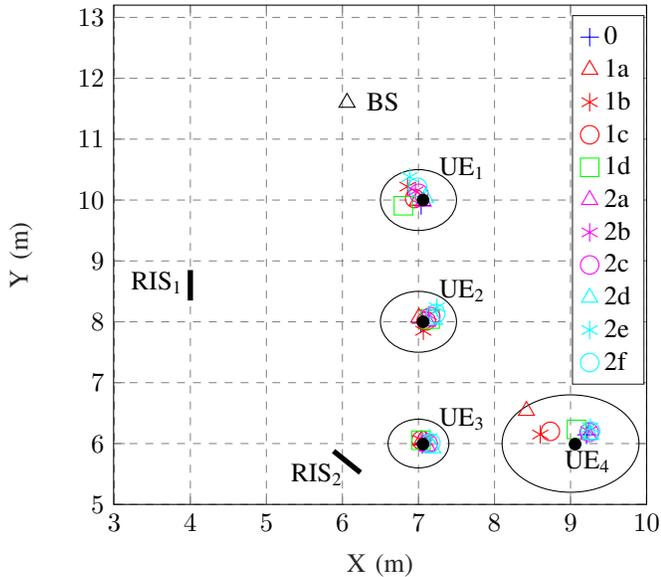
\begin{figure}[t]
  {
%
%
\definecolor{mycolor1}{rgb}{1.00000,0.00000,1.00000}%
\definecolor{mycolor2}{rgb}{0.00000,1.00000,1.00000}%
\begin{tikzpicture}

\begin{axis}[%
width=0.8\columnwidth,
height=0.75\columnwidth,
at={(2.237in,2.691in)},
scale only axis,
xmin=2.99109527027665,
xmax=10,
xlabel style={font=\normalsize \color{white!15!black}},
xlabel={X (m)},
ymin=5,
ymax=13.2012000852459,
ytick = {5,...,13},
ylabel style={font=\normalsize \color{white!15!black}},
ylabel={Y (m)},
axis background/.style={fill=white},
legend columns=1,
legend style={font = \normalsize, at={(0.86,0.61)}, anchor=west, legend cell align=left, align=left, draw=white!15!black},
xmajorgrids,
ymajorgrids,
major grid style={line width=.005pt,draw=black!50, dashed},
]

\addplot [color=blue, only marks, mark=+, mark size=3.5pt, mark options={solid, blue}]
  table[row sep=crcr]{%
7.035	9.91125046262034\\
7.03057141913445	7.97777815141599\\
7.0497946127015	5.98659580783041\\
9.21	6.14403995615804\\
};
\addlegendentry{0}

\addplot [color=red, only marks, mark size=3.5pt, mark=triangle, mark options={solid, red}]
  table[row sep=crcr]{%
6.9876950306324	9.99318507301552\\
7.00586468006825	8.06998495665315\\
7.03855276473988	6.05035149579406\\
8.42002592978546	6.53890805998185\\
};
\addlegendentry{1a}

\addplot [color=red, only marks, mark size=3.5pt, mark=asterisk, mark options={solid, red}]
  table[row sep=crcr]{%
6.85693839291954	10.2196622134373\\
7.06398911624905	7.85306160791237\\
7.03758965944646	6.05581353736839\\
8.60218968315809	6.14825663121947\\
};
\addlegendentry{1b}

\addplot [color=red, only marks, mark size=3.5pt, mark=o, mark options={solid, red}]
  table[row sep=crcr]{%
6.94973861297612	10.0255277601301\\
8.73427123282468	6.20042173942\\
7.10241603102591	8.01469884471163\\
7.03526868417252	6.05490310650092\\
};
\addlegendentry{1c}

\addplot [color=green, only marks, mark size=3.5pt, mark=square, mark options={solid, green}]
  table[row sep=crcr]{%
6.80171238051149	9.90645993898018\\
    7.15304091869556	8.04403381470423\\
7.03372130532332	6.05440557160854\\
9.07789867348452	6.23213696123717\\
};
\addlegendentry{1d}

\addplot [color=mycolor1, only marks, mark size=3.5pt, mark=triangle, mark options={solid, mycolor1}]
  table[row sep=crcr]{%
7.06312829768132	9.97636354317598\\
7.12092822030239	8.00577097970282\\
7.11858546144056	5.95603476105265\\
9.23819764211657	6.14413975212434\\
};
\addlegendentry{2a}

\addplot [color=mycolor1, only marks, mark size=3.5pt, mark=asterisk, mark options={solid, mycolor1}]
  table[row sep=crcr]{%
6.95991442211364	10.144543644714\\
7.13695127600421	8.10771997755953\\
7.09522056330512	6.03995948635316\\
9.23531259864484	6.20419826678718\\
};
\addlegendentry{2b}

\addplot [color=mycolor1, only marks, mark size=3.5pt, mark=o, mark options={solid, mycolor1}]
  table[row sep=crcr]{%
7.00369514647282	10.1101879710533\\
7.16172919115958	8.08306792103016\\
7.12593914559502	6.00179756220452\\
9.24567349384094	6.18421202722183\\
};
\addlegendentry{2c}

\addplot [color=mycolor2, only marks, mark size=3.5pt, mark=triangle, mark options={solid, mycolor2}]
  table[row sep=crcr]{%
7.09125659534519	10.0414766237316\\
7.21128502147032	8.03376380796982\\
7.18737631017774	5.92547371370666\\
9.26639528423314	6.14423954809071\\
};
\addlegendentry{2d}

\addplot [color=mycolor2, only marks, mark size=3.5pt, mark=asterisk, mark options={solid, mycolor2}]
  table[row sep=crcr]{%
6.88482884391792	10.377836826808\\
7.24333113287396	8.23766180370262\\
7.1406465139038	6.09332316482452\\
9.26062519728968	6.26435657741752\\
};
\addlegendentry{2e}

\addplot [color=mycolor2, only marks, mark size=3.5pt, mark=o, mark options={solid, mycolor2}]
  table[row sep=crcr]{%
6.98804271970953	10.2096567252618\\
7.22730807717214	8.13571280583627\\
7.16401141204086	6.00939843933282\\
9.26351024076141	6.20429806275352\\
};
\addlegendentry{2f}
\addplot [color=black, only marks, mark size=3.5pt, mark=triangle, mark options={solid, black}]
  table[row sep=crcr]{%
6.06	11.6\\
};

\addplot [color=black, line width=2.0pt,  forget plot]
  table[row sep=crcr]{%
4	8.35\\
4	8.85\\
};

\addplot [color=black, line width=2.0pt]
  table[row sep=crcr]{%
5.88322330470336	5.87677669529664\\
6.23677669529664	5.52322330470336\\
};

\addplot [color=black, only marks, mark size=2.2pt, mark=*, mark options={solid, black}]
  table[row sep=crcr]{%
7.06	10\\
7.06	8\\
7.06	5.99\\
9.06	5.99\\
};
\draw [black] (axis cs:7,10) ellipse [x radius=50, y radius=50];
\draw [black] (axis cs:7,8) ellipse [x radius=50, y radius=50];
\draw [black] (axis cs:7,6) ellipse [x radius=40, y radius=40];
\draw [black] (axis cs:9,6) ellipse [x radius=90, y radius=80];
\node[right, align=left]
at (axis cs:6.198,11.614) {BS};
\node[right, align=left]
at (axis cs:7.15,10.52) {$\text{UE}_\text{1}$};
\node[right, align=left]
at (axis cs:7.15,8.52) {$\text{UE}_\text{2}$};
\node[right, align=left]
at (axis cs:7.15,6.4) {$\text{UE}_\text{3}$};
\node[right, align=left]
at (axis cs:8.8,5.7) {$\text{UE}_\text{4}$};
\node[right, align=left]
at (axis cs:3.1,8.643) {$\text{RIS}_\text{1}$};
\node[right, align=left]
at (axis cs:5.2,5.498) {$\text{RIS}_\text{2}$};
\end{axis}

\end{tikzpicture}%
}
  \caption{Ground-truth and estimated \ac{UE} locations (over the 4 first tested locations) for the scenarios of Table \ref{Table:Positioning_Scenarios}.}
\label{fig:ex_pos_heatmap}
\end{figure}

Fig.~\ref{fig:ex_pos_heatmap} illustrates, qualitatively, the positioning results for 4 of the 5 tested \ac{UE} locations and the 11 scenarios of Table~\ref{Table:Positioning_Scenarios}, while Table~\ref{Table:Positioning_Results} reports, quantitatively, the corresponding errors per \ac{UE} location, along with the sample median error and \ac{RMSE} (both calculated over the 5 \ac{UE} tested locations). In this table, the symbol ↓ indicates that the error is significantly higher than the best result, typically according to Wilcoxon's rank sum test used at the \emph{p}-value threshold of 0.01 \cite{wilcoxon}. The latter entries correspond mostly to test configurations where the RP1 is used at large \ac{RIS}-\ac{UE} distances, typically with respect to UE5 locations and, to a minor extent, with respect to UE4. First, this is in line with the link-level radio parameters estimation results in Sec. \ref{subsec:multipath_parameters_estimation}, where \ac{AoD} estimation errors can typically exceed $10\degree$ as \ac{RIS}-\ac{UE} distances reach $5$m-$6$m, whatever the estimation method. Beyond, regardless of the quality of radio parameters estimation, due to the $5\degree$ \ac{AoD} discretization step used for beam sweeping in azimuth, positioning errors naturally tend to be larger at large distances, due to obvious geometric dilution considerations.   

Nevertheless, it is also observed that despite the use of poorly informative radiolocation metrics such as that related to received power (herein, \ac{BS} and \ac{RIS} \acp{AoD} based on overall channel gain), single-\ac{BS} still seems feasible in Scenarios 1a to 1c within an accuracy level comparable to that of the reference scenario 0. 
Mitigating a little bit the previous result, it should be recalled that channel sounding equipment usually benefits from better signal dynamics and sensibility than that of integrated receivers at real \ac{UE} terminals. Accordingly, the latter may be subject to relatively large fluctuations of the received power (in comparison with the power gain brought by \ac{RIS} reflections), hence making the detection of \ac{RIS} \acp{AoD} more challenging in practice. 

Finally, as regards to the expected localization boost, none of the tested RIS-aided configurations (in terms of both radio metrics or \ac{UE} locations) could really outperform this reference scenario, suggesting that passive \ac{RRIS} are most likely beneficial to enable non-feasible localization configurations, rather than improving localization accuracy.

%
%

\begin{table}
\renewcommand{\arraystretch}{1.1}
\centering
\scriptsize
\caption{LS positioning errors (in m) for the scenarios of Table \ref{Table:Positioning_Scenarios}.}
\begin{threeparttable}
\begin{tabular}{|c|c|c|c|c|c|c|c|}
\hline
\textbf{Scenario} & \textbf{UE1} & \textbf{UE2} & \textbf{UE3} & \textbf{UE4} & \textbf{UE5} & \textbf{RMSE} & \textbf{Median} \\ 
\hhline{|========|}
0 & 0.09 & 0.04 & 0.01 & 0.21 & 0.22 & 0.15 & 0.09 \\ \hline 
1a & 0.07 & 0.09 & 0.06 & \textbf{0.84↓} & \textbf{7.16↓} & \textbf{3.22↓} & 0.08 \\ \hline 
1b & 0.30 & 0.15 & 0.07 & 0.48 & \textbf{0.83↓} & 0.46 & 0.30 \\ \hline 
1c & 0.11 &  0.05 & 0.07 & 0.39 & \textbf{2.99↓} & \textbf{1.35↓} & 0.11 \\ \hline 
1d & 0.27 & 0.10 & 0.07 & 0.24 & \textbf{6.78↓} & \textbf{3.04↓} & 0.24 \\ \hline 
2a & 0.02 & 0.06 & 0.07 & 0.24 & \textbf{1.40↓} & \textbf{0.64↓} & 0.07 \\ \hline 
2b & 0.18 & 0.13 & 0.06 & 0.28 & 0.44 & 0.25 & 0.18 \\ \hline 
2c & 0.12 & 0.13 & 0.07 & 0.27 & \textbf{0.97↓} & 0.46 & 0.13 \\ \hline 
2d & 0.05 & 0.15 & 0.14 & 0.26 & \textbf{2.60↓} & \textbf{1.17↓} & 0.16 \\ \hline 
2e & 0.42 & 0.30 & 0.13 & 0.34 & 0.76 & 0.44 & 0.34 \\ \hline 
2f &  0.22 & 0.22 & 0.11 & 0.30 & \textbf{1.35↓} & \textbf{0.63↓} & 0.22 \\ \hline 
\end{tabular}
\end{threeparttable} \label{Table:Positioning_Results}
\end{table}


%
%
\section{Conclusion and Future Work}
In this paper, we account for first 
experimental validations of \ac{RIS}-aided single-\ac{BS} positioning based on real frequency domain mmWave channel sounding measurements. Basic \ac{LS} positioning results show that the \acp{AoD} of 2 \ac{RIS}-reflected paths (or even the \ac{AoD} and delay of 1 single \ac{RIS}-reflected path) could viably (i) replace the missing direct path in case of \ac{NLoS} situation between the \ac{BS} and the \ac{UE} or (ii) complete this direct path in \ac{LoS}, both given that the \ac{RIS}-\ac{UE} distance remains on the order of a few meters at most. Beyond, it is noted that the estimation of \ac{RIS}-reflected path parameters could be significantly degraded, thus making counterproductive \ac{RIS} contribution to the final positioning result.

Leveraging the same measurement data, future works will investigate the impact of bandwidth occupancy on performance, as well as the suppression of the systematic static multipath components to ease the detection of \ac{RIS}-reflected paths, hence getting rid of the genie-aided spatial pre-filtering step in the Delay-\ac{AoA} domain. Other possible \ac{RIS}-based applications in the same context concern the localization of passive objects for opportunistic mapping purposes. 

\section*{Acknowledgment}
This work has been supported, in part, by the EU H2020 RISE-6G project under grant 101017011. 
The authors would also like to warmly thank J.-B. Gros from Greenerwave, for providing the \ac{RRIS} prototype involved in the measurement campaign reported here, as well as for his technical assistance and meaningful advice.

\balance
\bibliographystyle{ieeetr}
\bibliography{references}

\end{document}